\documentclass[aps,pra,twocolumn,amsmath,
showpacs]{revtex4-1}
\usepackage{epsfig}
\begin{document}
\title{Gradient corrections to the local density approximation \\
for trapped superfluid Fermi gases}
\author{Andr\'as Csord\'as}
\email{csordas@tristan.elte.hu}
\affiliation{HAS-ELTE Statistical and Biological Physics Research Group,
             P\'azm\'any P. S\'et\'any 1/A, H-1117 Budapest, Hungary}
\affiliation{Department of Physics of Complex Systems, E\"otv\"os University,
             P\'azm\'any P. S\'et\'any 1/A, H-1117 Budapest, Hungary}
\author{Orsolya Alm\'asy}
\email{oalmasy@gmail.com}
\affiliation{Institute for Physics, E\"otv\"os University,
             P\'azm\'any P. S\'et\'any 1/A, H-1117 Budapest, Hungary} 
\altaffiliation{present address: Semilab Semiconductor Physics
  Laboratory Co. Ltd., 
             Prielle Korn\'elia u. 2. H-1117 Budapest, Hungary}
\author{P\'eter Sz\'epfalusy}
\email{psz@complex.elte.hu}
\affiliation{Department of Physics of Complex Systems, E\"otv\"os University,
             P\'azm\'any P. S\'et\'any 1/A, H-1117 Budapest, Hungary}
\affiliation{Research Institute for Solid State Physics and Optics,
             P. O. Box 49, H-1525 Budapest, Hungary}
\date{\today}
\begin{abstract}
Two species superfluid Fermi gas is investigated on the BCS side
up to the Feshbach resonance. Using the Greens's function technique
gradient corrections are calculated to the generalized Thomas-Fermi
theory including Cooper pairing. Their relative magnitude is found
to be measured by the small parameter $(d/R_{TF})^4$, where $d$ is the
oscillator length of the trap potential and $R_{TF}$ is the radial
extension of the density $n$ in the Thomas-Fermi approximation. In particular
at the Feshbach resonance the universal 
corrections to the local density approximation are calculated and a
universal prefactor $\kappa_W=7/27$ is derived for the von Weizs\"acker
type correction $\kappa_W(\hbar^2/2m)(\nabla^2 n^{1/2}/n^{1/2})$.
\end{abstract}
\maketitle

\section{Introduction}
\label{sec:intro}

Fermi gases below the degeneracy temperature have been the subject of 
intensive research in the last years both experimentally and
theoretically (See for reviews \cite{Bloch08,Giorgini08}). 
Particular interest has been devoted to the possible superfluid state 
whose creation and properties have been studied for 
both negative and positive values of   $s$-wave scattering
lengths $a$, characterizing the interaction between the particles. 
At the Feshbach resonance \cite{Feshbach68,Tiesinga92,Inouye98,Regal03,Chin10} 
$a$ becomes infinity and 
certain universal behavior shows up.
An important aspect 
of the problem is that the gas is trapped and thereby is inhomogeneous.
When the energy gap function exceeds the level spacing near the Fermi sea
a local density approximation (LDA) is applicable. 
As a simplest approach in its spirit neglecting the space
gradients of the density and the gap function the Thomas-Fermi theory was
generalized to include superfluid pair correlation results \cite{Szepfalusy64},
when the system is treated in the generalized Hartree-Fock method 
\cite{Valatin61}. Since the Thomas-Fermi approximation is widely
used in case of trapped gases it is desirable to investigate systematically
the corrections to it, even if they are expected to be small 
for large particle numbers, except in the surface region 
(Here the gradient corrections make
explicitely visible the limits of the usual LDA results).
For particle
numbers, however, which are treated in Monte-Carlo simulations the
Thomas-Fermi theory needs corrections. More importantly it makes
possible to extend the concept of universality at unitarity
\cite{csordas07}. In particlular we derive in this paper a universal
prefactor for the von Weizs\"acker type correction to the generalized
Thomas-Fermi theory (see for a review of the von Weizs\"acker correction
in normal systems \cite{Parr89}).

In the present paper gradient
corrections are calculated up to second order at zero temperature. 
Baranov \cite{Baranov99} studied 
the gradient corrections even at finite temperatures in cases when
Eilenberger's equations \cite{Eilenberger68} are applicable. That  approach
is different from ours, which is free from this restriction. 
The applied technique here is based upon
the equation of motion as expressed in terms of the Green's functions.
The method has been developed first to the electron gas of the atoms,
which is of course a normal system \cite{Baraff61}. 
It has been generalized to superfluid
state somewhat later independently for superconductors 
in slowly varying magnetic field \cite{Werthamer63} and 
for nuclei \cite{Szepfalusy64}. The
latter work is most closely related to the present one. 
The resulting expressions are rather cumbersome, but considerably
simplify at unitarity.
To evaluate them we choose the mean-field BCS (MF-BCS) model introduced
by Leggett, Eagles, Nozi\`eres and Schmitt-Rink
\cite{leggett80,eagles69,nozieres85},  
which neglects the
self-consistent Hartree-type terms. We start however, from the generalized
Hartree-Fock (GHF) model  \cite{Valatin61} to present the results in a
more complete form for future use. The Hamiltonian is 
\begin{gather}
H=\sum_\sigma\int d^3r \,
\psi^+_\sigma(\mathbf{r})\left(-\frac{\hbar^2}{2m}\nabla^2+
U_{ext}(\mathbf{r})-
\mu\right)\psi_\sigma(\mathbf{r}) \nonumber\\
+\frac{1}{2}\sum_{\sigma,\sigma'}\int d^3r d^3 r' \,
\psi^+_\sigma(\mathbf{r})
\psi^+_{\sigma'}(\mathbf{r}') v(\mathbf{r},\mathbf{r}')
{\psi_{\sigma'}}(\mathbf{r}')\psi_\sigma(\mathbf{r}),
\label{eq:horig}
\end{gather}
where $U_{ext}(\mathbf{r})$ is the trapping potential,
$\mu$ is the chemical potential, $v(\mathbf{r}-\mathbf{r}')$
describes the interaction and
$\sigma$ stands for the internal degrees of freedoms.
We assume two equally populated hyperfine states and 
$\sigma=\uparrow,\downarrow$ 
will be termed as spin.
In GHF approximation the Hamiltonian simplifies to
\begin{gather}
H_\textrm{mf}=\sum_\sigma\int d^3r \,
\psi^+_\sigma(\mathbf{r})\left(-\frac{\hbar^2}{2m}\nabla^2+U_{ext}(\mathbf{r})-
\mu\right)\psi_\sigma(\mathbf{r}) \nonumber\\
+\sum_{\sigma,\sigma'}\int \!\! d^3r\, d^3 r' \,
v(\mathbf{r},\mathbf{r}') 
h_{\sigma,\sigma}(\mathbf{r},\mathbf{r})
\psi^+_{\sigma'}(\mathbf{r}')\psi_{\sigma'}(\mathbf{r}')
\nonumber\\
-\sum_{\sigma,\sigma'}\int \!\! d^3r\, d^3 r' \,
v(\mathbf{r},\mathbf{r}') 
h_{\sigma,\sigma'}(\mathbf{r},\mathbf{r}')
\psi^+_{\sigma}(\mathbf{r})\psi_{\sigma'}(\mathbf{r}')
\nonumber\\
+\frac{1}{2}\sum_{\sigma,\sigma'}\int \!\! d^3r\, d^3 r' \,
v(\mathbf{r},\mathbf{r}')\! \left(
\chi_{\sigma,\sigma'}(\mathbf{r},\mathbf{r'})
\psi^+_\sigma(\mathbf{r})\psi^+_{\sigma'}(\mathbf{r'}) 
+\textrm{Hc}
\right)\! . \label{eq:mfham}
\end{gather}
Here
\begin{eqnarray}
n(\mathbf{r})_\textrm{tot}&=&\sum_\sigma n_\sigma(\mathbf{r})=
2n(\mathbf{r}), \nonumber\\
n_\sigma(\mathbf{r})&=& h_{\sigma,\sigma}(\mathbf{r},\mathbf{r}),
\nonumber\\
h_{\sigma',\sigma}(\mathbf{r}',\mathbf{r})&=&\left<\psi^+_\sigma(\mathbf{r})
\psi_{\sigma'}(\mathbf{r}')\right>.
\label{eq:denskordef}
\end{eqnarray}
The first line in (\ref{eq:mfham}) contains the one-particle term of 
(\ref{eq:horig}),
the second line is the Hartree term, the third is the Fock term and
furthermore the Cooper pairing is represented by the last line, where
\begin{equation}
\chi_{\sigma',\sigma}(\mathbf{r}',\mathbf{r})=
\left<\psi_\sigma(\mathbf{r})\psi_{\sigma'}(\mathbf{r}')\right> .
\label{eq:chidef}
\end{equation}
The correlation functions $\chi$ and $h$ have to be 
determined self-consistently. 

We shall consider the special case when the interaction
can be approximated by a contact potential
\begin{equation}
v(\mathbf{r},\mathbf{r}')=\frac{4\pi\hbar^2a}{m}
\delta(\mathbf{r}-\mathbf{r}')\equiv g\delta(\mathbf{r}-\mathbf{r}').
\label{eq:contactinteraction}
\end{equation}
In case of contact interaction the first three lines of the
Hamiltonian (\ref{eq:mfham}) 
can safely  joined together as follows
\begin{gather}
H_\textrm{mf}=\sum_\sigma\int d^3r \,
\psi^+_\sigma(\mathbf{r})\left(-\frac{\hbar^2}{2m}\nabla^2+U(\mathbf{r})-
\mu\right)\psi_\sigma(\mathbf{r}) \nonumber\\
+\frac{1}{2}\sum_{\sigma,\sigma'}\int \!\! d^3r\, d^3 r' \,
v(\mathbf{r},\mathbf{r}')\! \left(
\chi_{\sigma,\sigma'}(\mathbf{r},\mathbf{r'})
\psi^+_\sigma(\mathbf{r})\psi^+_{\sigma'}(\mathbf{r'}) 
+\textrm{Hc}
\right)\!, \label{eq:mfham1}
\end{gather}
where 
\begin{equation}
U(\mathbf{r})=U_{ext}(\mathbf{r})+ gn(\mathbf{r}),
\end{equation}
but we keep the fourth line of (\ref{eq:mfham}) as it is,
because  $\chi_{\sigma,\sigma'}(\mathbf{r},\mathbf{r})$ is not
a well-defined object.

The paper is organized as follows. In Section \ref{sec:formulation}. we
present the equations for the Green's functions, while in Section
\ref{sec:pert_series}. their perturbation series are presented. The
self-consistent scheme for the density and the gap function is worked
out in Section \ref{sec:gradexpofphysq}. up to second order in $\hbar$
to the local density approximation, which can be regarded as a
generalized Thomas-Fermi theory. In the second part of the paper the
MF-BCS model is applied. In Section \ref{sec:secordercorr}. the second
order corrections are evaluated perturbatively in the case of a
general external potential. Section \ref{sec:univprefactor}. is devoted
to the problems of the unitary gas, in particular the prefactor of the
von Weizs\"acker type correction is calculated. In Section
\ref{sec:isotropictrap}. the trap potential is assumed to be an
isotropic harmonic one to make some features more visible.  Section
\ref{sec:summary}. contains the discussion of the results.

\section{Formulation}
\label{sec:formulation}

The gradient expansion can be best derived using the one particle normal
\begin{equation}
\label{eq:greendef}
G_{\sigma,\sigma'}(\mathbf{r}_1,t_1;\mathbf{r}_2,t_2)=
-i\left<T\psi_{\sigma}(\mathbf{r}_1,t_1)
\psi^+_{\sigma'}(\mathbf{r}_2,t_2)\right>
\end{equation}
and the anomalous
\begin{equation}
\label{eq:anomdef}
F_{\sigma,\sigma'}(\mathbf{r}_1,t_1;\mathbf{r}_2,t_2)=
-i\left<T\psi^+_{\sigma}(\mathbf{r}_1,t_1)\psi^+_{\sigma'}(\mathbf{r}_2,t_2)
\right>,
\end{equation} 
Green's functions \cite{Werthamer63,Szepfalusy64}.

If the Hamiltonian is time independent, which is the case
we want to discuss,  the Green's functions depend
on the combination $t_1-t_2$ not separately on $t_1$ and $t_2$ (i.e.,
$G_{\sigma,\sigma'}(\mathbf{r}_1,t_1;\mathbf{r}_2,t_2)\equiv
G_{\sigma,\sigma'}(\mathbf{r}_1,\mathbf{r}_2,t_1-t_2)$, and similarly for $F$).
Correlation functions (\ref{eq:chidef}) and (\ref{eq:denskordef}) can
be calculated from $G$ and $F$ by the limiting procedures
\begin{eqnarray}
h_{\sigma',\sigma}(\mathbf{r}',\mathbf{r})&=&-i \lim_{\varepsilon \to 0}
G_{\sigma',\sigma}(\mathbf{r}',\mathbf{r},\varepsilon) 
\label{eq:hdef1}\\ 
\chi_{\sigma',\sigma}(\mathbf{r}',\mathbf{r})&=&-i \lim_{\varepsilon \to 0}
F^*_{\sigma',\sigma}(\mathbf{r}',\mathbf{r},-\varepsilon).
\label{eq:chidef1}
\end{eqnarray}
We consider the problem of singlet Cooper paring. In that case
the  nonvanishing elements of the Green's functions can be chosen to be
$G_{\uparrow\uparrow}=G_{\downarrow\downarrow}$ and 
$F_{\uparrow\downarrow}=-F_{\downarrow\uparrow}$ respectively \cite{Landau}. 
For practical purposes let us introduce the functions
\begin{gather}
\nu(\mathbf{r}_1,\mathbf{r}_2)=
\left(-\frac{\hbar^2}{2m}\nabla^2_1+U(\mathbf{r}_1)-
\mu\right)\delta(\mathbf{r}_1-\mathbf{r}_2),
\label{eq:nudef}\\
\Delta_{\sigma\sigma'}(\mathbf{r}_1,\mathbf{r}_2)=
v(\mathbf{r}_1-\mathbf{r}_2)\chi_{\sigma\sigma'}
(\mathbf{r}_1,\mathbf{r}_2).
\label{eq:mudef}
\end{gather}
Then, the time evolutions of the two Green's functions can be written as
\begin{gather}
i\hbar\frac{\partial}{\partial
  t_1}G_{\uparrow\uparrow}(\mathbf{r}_1,\mathbf{r}_2,t_1-t_2)=
\hbar\delta(t_1-t_2)\delta(\mathbf{r}_1-\mathbf{r}_2)+\nonumber\\
+\int d^3r\,
\nu(\mathbf{r}_1,\mathbf{r})G_{\uparrow\uparrow}
(\mathbf{r},\mathbf{r}_2,t_1-t_2)\nonumber\\
+\int d^3r\, \Delta_{\uparrow\downarrow}(\mathbf{r}_1,\mathbf{r})
F_{\uparrow\downarrow}(\mathbf{r},\mathbf{r}_2,t_1-t_2),
\label{eq:gemrt}
\end{gather}
and
\begin{gather}
i\hbar\frac{\partial}{\partial t_1}F_{\downarrow\uparrow}
(\mathbf{r}_1,\mathbf{r}_2,t_1-t_2)= \nonumber\\
-\int d^3r\,\Delta^*_{\downarrow\uparrow}(\mathbf{r}_1,\mathbf{r})
G_{\uparrow\uparrow}(\mathbf{r},\mathbf{r}_2,t_1-t_2) \nonumber\\
-\int d^3r\,\nu(\mathbf{r}_1,\mathbf{r})
F_{\downarrow\uparrow}(\mathbf{r},\mathbf{r}_2,t_1-t_2).
\label{eq:femrt}
\end{gather}
The symbol '$*$' denotes complex conjugation.
Let us take the Fourier transform with respect to time
of the Green's functions  as
\begin{equation}
G_{\uparrow\uparrow}(\mathbf{r}_1,\mathbf{r}_2,\omega)=
\int_{-\infty}^\infty dt\, e^{i\omega t}
G_{\uparrow\uparrow}(\mathbf{r}_1,\mathbf{r}_2,t)
\label{eq:fouriertwrt}
\end{equation}
(and similarly for $F$). Next we transform quantities like 
$A(\mathbf{r}_1,\mathbf{r}_2)$ in Eqs. 
(\ref{eq:gemrt}) and (\ref{eq:femrt}) to mixed position-momentum
representation by introducing $\mathbf{R}=(\mathbf{r}_1+\mathbf{r}_2)/2$
and $\mathbf{r}=\mathbf{r}_1-\mathbf{r}_2$ and taking the Fourier
transform with respect to $\mathbf{r}$:
\begin{equation}
A(\mathbf{R},\mathbf{p})=\int d^3k\, 
e^{i\mathbf{p}\mathbf{r}/\hbar}
A\left(\mathbf{R}+\mathbf{r}/2, \mathbf{R}-\mathbf{r}/2\right).
\end{equation}
We use the term phase space for the $(\mathbf{R},\mathbf{p})$ space
in the following.
If a quantity $C(\mathbf{r}_1,\mathbf{r}_2)$ is given by
\begin{equation}
C(\mathbf{r}_1,\mathbf{r}_2)=\int d^3 r
A(\mathbf{r}_1,\mathbf{r})B(\mathbf{r},\mathbf{r}_2)
\end{equation}
then Baraff and Borowitz \cite{Baraff61,Baraff61a} showed 
that the corresponding relation
in position-momentum space can be expressed as
\begin{equation}
C(\mathbf{R},\mathbf{p})=\Theta\left[ 
A(\mathbf{R},\mathbf{p}), B(\mathbf{R},\mathbf{p})\right],
\end{equation}
where $\Theta$ is a bilinear operator acting  
on two phase space functions as \cite{Groenewold46}
\begin{gather}
\Theta\left[ 
A(\mathbf{R},\mathbf{p}), B(\mathbf{R},\mathbf{p})\right]=
\lim_{\begin{array}{c}
\mathbf{R}' \to \mathbf{R}
\\
\mathbf{p}' \to \mathbf{p}
\end{array}}
\nonumber \\
\exp\left[ 
\frac{i\hbar}{2}\sum_{i=1}^3 \left( 
\frac{\partial}{\partial R_i}\frac{\partial}{\partial
      p'_i}-\frac{\partial}{\partial R'_i}\frac{\partial}{\partial p_i}
\right)\right]A(\mathbf{R},\mathbf{p})B(\mathbf{R}',\mathbf{p}').
\label{eq:theta_def}
\end{gather}
Eqs. (\ref{eq:gemrt}) and (\ref{eq:femrt}) 
in the $\mathbf{R},\mathbf{p},\omega$ representation 
can be written in the compact forms
\begin{eqnarray}
\hbar &=& \hbar\omega G -\Theta[\nu, G]- \Theta[\Delta, F], \label{eq:Geqorig}\\
0     &=& \hbar\omega F +\Theta[\nu, F]- \Theta[\Delta^*, G],\label{eq:Feqorig}
\end{eqnarray}
where $G \equiv G_{\uparrow\uparrow}(\mathbf{R},\mathbf{p},\omega)$, 
$F \equiv F_{\downarrow\uparrow}(\mathbf{R},\mathbf{p},\omega)$, 
$\Delta\equiv\Delta_{\uparrow\downarrow}(\mathbf{R},\mathbf{p})$, 
$\nu\equiv\nu(\mathbf{R},\mathbf{p})$, respectively.
In deriving Eqs. (\ref{eq:Geqorig}), (\ref{eq:Feqorig}) we used the 
properties
\begin{eqnarray}
\Delta_{\sigma\sigma'}(\mathbf{r}_1,\mathbf{r}_2)&=&
-\Delta_{\sigma'\sigma}(\mathbf{r}_2,\mathbf{r}_1)\nonumber\\
\Delta_{\downarrow\uparrow}(\mathbf{R},-\mathbf{p})&=&
\Delta_{\uparrow\downarrow}(\mathbf{R},\mathbf{p})\nonumber
\end{eqnarray}
which can be proven from the definition of $\Delta$ (Eq. (\ref{eq:mudef})).
The $\omega$ independent functions $\nu$ and $\Delta$ in the
mixed representation are
\begin{equation}
\nu(\mathbf{R},\mathbf{p})=\frac{p^2}{2m}+U(\mathbf{R})-\mu
\label{eq:nurplambda}
\end{equation}
and 
\begin{equation}
\Delta(\mathbf{R},\mathbf{p})=
\int d^3r\, e^{-i\mathbf{pr}/\hbar}
v(\mathbf{r}) 
\chi(\mathbf{R}+\mathbf{r}/2,\mathbf{R}-\mathbf{r}/2).
\label{eq:gapcalcfromchi}
\end{equation}

Green's function are useful for calculating physical quantities
such as the density $n(\mathbf{R})\equiv n_\uparrow(\mathbf{R})$,
or the equal-time expectation values  $h(\mathbf{R},\mathbf{p})\equiv
h_{\uparrow\uparrow}(\mathbf{R},\mathbf{p})$ and  
$\chi(\mathbf{R},\mathbf{p})\equiv 
\chi_{\uparrow\downarrow}(\mathbf{R},\mathbf{p})$
(defined in Eqs. (\ref{eq:denskordef}) 
and (\ref{eq:chidef})). From Eqs. (\ref{eq:hdef1}) and 
(\ref{eq:chidef1}) follow that
\begin{equation}
n(\mathbf{R})=\int \!\!\frac{d^3p}{(2\pi\hbar)^3}h(\mathbf{R},\mathbf{p}),
\label{eq:ncalc}
\end{equation}
\begin{equation}
h(\mathbf{R},\mathbf{p})=-i\int \frac{d\omega}{2\pi}
G(\mathbf{R},\mathbf{p},\omega)e^{i\omega\varepsilon},
\label{eq:hcalfromG}
\end{equation}
\begin{equation}
\chi(\mathbf{R},\mathbf{p})^*=-i\int \!\frac{d\omega}{2\pi}
F(\mathbf{R},\mathbf{p},\omega)e^{i\omega\varepsilon},
\label{eq:chicalfromF}
\end{equation}
where $\varepsilon$ is an infinitesimally small positive regularization
parameter. 

The widely used interaction potential (\ref{eq:contactinteraction}) 
leads to divergence in the gap equation, which requires
some special care. Due to the $\delta$ interaction
$\Delta(\mathbf{R},\mathbf{p})$ is momentum independent. In that case,
the self-consistent equation for the local gap
$\Delta(\mathbf{R})$ has to be regularized. It means, that 
we should take the regularized part $F_{reg}$ of $F$ by which the 
the self consistent gap-equation 
\begin{equation}
\Delta(\mathbf{R})^*=\frac{4\pi\hbar^2 a}{im}
\int \!\!\frac{d^3p}{(2\pi\hbar)^3}\int\frac{d\omega}{2\pi}
F_{reg}(\mathbf{R},\mathbf{p},\omega)e^{i\omega\varepsilon}
\end{equation}
provides a finite value for $\Delta(\mathbf{R})$ \cite{stoof96,bruun99}.

\section{Perturbation series for $G$ and $F$}
\label{sec:pert_series}

In this section we shall construct a formal solution
of Eqs. (\ref{eq:Geqorig}) and (\ref{eq:Feqorig}) supposing that
the functions $\nu$ and $\Delta$ are known.
The bilinear operator $\Theta$ as defined in Eq. (\ref{eq:theta_def}) 
can be expanded as a formal series of $\hbar$:
\begin{equation}
\Theta \left[ K_1,K_2 \right] =\sum_{j=0}^\infty {\hbar}^j
\Theta_j\left[ K_1,K_2\right]. 
\label{eq:theta_series}
\end{equation}
The first two operators $\Theta_0$ and $\Theta_1$
are simply
\begin{eqnarray}
\Theta_0\left[ K_1,K_2 \right]&=&K_1\cdot K_2, \nonumber \\
\Theta_1\left[ K_1,K_2 \right]&=&\frac{i}{2}\{K_1,K_2 \},
\end{eqnarray}
where $\{\ldots\}$ is a usual Poisson bracket. For higher order
terms in the series (\ref{eq:theta_series}) it is useful to treat
derivatives according to the phase space variables
on equal footing by the definition
\begin{equation}
\{ {\partial}_i
\}_{i=1}^6 \equiv
\left(\frac{\partial}{\partial R_1},\frac{\partial}{\partial
  R_2},\frac{\partial}{\partial R_3},\frac{\partial}{\partial
  p_1},\frac{\partial}{\partial p_2},
\frac{\partial}{\partial p_3}\right). 
\end{equation}
We also need an antisymmetric metric $g^{\alpha\beta}$, where
$g^{14}=g^{25}=g^{36}=-g^{41}=-g^{52}=-g^{63}=1$ and all the other
elements are zero. The metrics reflect the simplectic structure
of the phase space. For example
\begin{equation} 
\Theta_2\left[ K_1,K_2 \right]=-\frac{1}{8}
g^{\alpha\beta}g^{\gamma\delta}
\left(\partial_\alpha\partial_\gamma K_1 \right)
\left(\partial_\beta \partial_\delta K_2 \right)
\end{equation}
Expressions for higher order $\Theta_j$'s 
can be derived similarly in a straightforward manner.
Let us write now the normal and the anomalous Green's functions
$G$ and $F$ as a formal power series in $\hbar$ 
\begin{eqnarray}
G(\mathbf{R},\mathbf{p},\omega)&=&\hbar\sum_{j=0}^\infty {\hbar}^j
G_j(\mathbf{R},\mathbf{p},\omega) 
\label{eq:Gformexp}\\
F(\mathbf{R},\mathbf{p},\omega)&=&\hbar\sum_{j=0}^\infty {\hbar}^j
F_j(\mathbf{R},\mathbf{p},\omega) 
\label{eq:Fformexp}
\end{eqnarray}
If we write 
\begin{equation}
\Omega=\hbar\omega
\end{equation}
and treat this quantity as an 
$o(\hbar^0)$ term then we get from (\ref{eq:Geqorig}) and (\ref{eq:Feqorig})
in different orders of $\hbar$ the following equations
\begin{eqnarray}
(\Omega-\nu)G_j-\Delta F_j&=&Q_j, \label{eq:firstperteq}\\ 
-{\Delta}^*G_j+(\Omega+\nu)F_j
&=&P_j, \label{eq:secndperteq}
\end{eqnarray}
with 
\begin{eqnarray}
Q_0&=&1,\label{eq:Q0def}\\
P_0&=&0,
\end{eqnarray}
and for $j\ge 1$
\begin{eqnarray}
   Q_j&=&\sum_{k=1}^j
\Big(\Theta_k\left[\nu, G_{j-k}\right]+\Theta_k \left[\Delta,
  F_{j-k}\right]\Big) \label{eq:Qj}\\  
P_j&=&\sum_{k=1}^j
\Big(\Theta_k \left[{\Delta}^*,G_{j-k}\right]
-\Theta_k\left[\nu, F_{j-k}\right]
\Big).
\label{eq:Pj}
\end{eqnarray}
It is clear from this structure that $Q_j$ and $P_j$ for fixed $j$
are given in terms of lower order corrections of $G$ and $P$.
Solutions to (\ref{eq:firstperteq}) and (\ref{eq:secndperteq})
are
\begin{eqnarray}
G_j&=&\frac{1}{{\Omega}^2-{E}^2}\left[(\Omega+\nu)Q_j+\Delta P_j\right]
\label{eq:grj}\\ 
F_j&=&\frac{1}{{\Omega}^2-{E}^2}\left[(\Omega-\nu)P_j+{\Delta}^\ast Q_j\right]
\label{eq:frj},
\end{eqnarray}
where
\begin{equation}
E=E(\mathbf{R},\mathbf{p})=\sqrt{\nu^2(\mathbf{R},\mathbf{p})+
|\Delta(\mathbf{R},\mathbf{p})|^2}
\label{eq:ERpdef}
\end{equation}
Using Equations (\ref{eq:Q0def})-(\ref{eq:ERpdef}) one can calculate
corrections to $G$ and $F$ up to arbitrary large orders.

Up to now, we have not addressed the question of the correct
pole structure of $G$ and $F$. This requires to introduce infinitesimal
imaginary parts in the denominators of $G$ and $F$.  
This step can be easily performed if we write the corrections as 
partial fractions in $\Omega$ with $\Omega$ independent numerators
and choose $i\delta$ accordingly to
\begin{eqnarray}
G_j&=&\sum_k
\left[\frac{A_{j,k}(\mathbf{R},\mathbf{p})}{(\hbar\omega-E+i\delta)^k}+
\frac{B_{j,k}(\mathbf{R},\mathbf{p})}{(\hbar\omega+E-i\delta)^k}\right],
\label{eq:grab}\\ 
F_j&=&\sum_k \left[\frac{C_{j,k}(\mathbf{R},\mathbf{p})}
{(\hbar\omega-E+i\delta)^k}+
\frac{D_{j,k}(\mathbf{R},\mathbf{p})}{(\hbar\omega+E-i\delta)^k}\right].
\label{eq:fcd}
\end{eqnarray}
The zeroth order coefficients are
\begin{equation}
A_{0,1}=\frac{1}{2}\left(1+\frac{\nu}{E} \right),\quad
B_{0,1}=\frac{1}{2}\left(1-\frac{\nu}{E} \right),
\label{eq:a01b01}
\end{equation}
and
\begin{equation}
C_{0,1}=-D_{0,1}=\frac{\Delta^*}{2E}
\label{eq:c01d01}
\end{equation}
All the other coefficients are zero.
Non-vanishing first order corrections involve 
Poisson brackets in the combinations of
\begin{eqnarray}
A_{1,1}&=&
-\frac{i}{8E^3}
\left(
\Delta^*
  \{\nu,\Delta\}-\Delta \{\nu,\Delta^*\}+\nu\{\Delta,\Delta^*\}
\right) \nonumber\\
&=&-B_{1,1},
\label{eq:a11b11}
\end{eqnarray}
\begin{equation}
A_{1,2}=E B_{1,1}+\frac{i}{8E}\{\Delta,\Delta^*\}, \,\,\,
B_{1,2}=E B_{1,1}-\frac{i}{8E}\{\Delta,\Delta^*\}.
\end{equation}
See also Ref. \cite{Szepfalusy64}.
It is important to note that for real $\Delta$ the 
first order correction $G_1$ to the normal Green's function
is identically zero. Coefficients of $F_1$ are nonzero
even for real $\Delta$ as can be seen from
\begin{equation}
C_{1,2}=-D_{1,2}=-\frac{i}{4E}\{\nu,\Delta^*\}.
\end{equation}
There are no first order poles of (\ref{eq:fcd}) for $j=1$, consequently, 
\begin{equation}
C_{1,1}=D_{1,1}=0.
\label{eq:c11d11}
\end{equation}
Higher than first order coefficients require tedious calculations.
Here we do not give explicitely the second order coefficient functions
in the numerators of Eqs. (\ref{eq:grab}), (\ref{eq:fcd}).
Instead, we sketch the structure of these corrections.
$G_2$ and $F_2$ involve $k=1,\ldots,4$ and the coefficient functions
for real $\Delta$
are linear combinations of ten (usual and) generalized Poisson brackets
$\{\Delta;\nu,\nu\}$, $\{\Delta,\Delta\}_+$, $\{\Delta;\Delta,\Delta\}$,  $\{\Delta;\nu,\Delta\}$,
$\{\nu;\Delta,\Delta\}$, $\{\nu,\Delta\}^2$, $\{\nu,\nu\}_+$, $\{\nu;\nu,\nu\}$,
$\{\nu;\Delta,\nu\}$, $\{\nu,\Delta\}_+$. The first
generalized Poisson bracket is
defined as
\begin{equation}
\{A,B\}_{+}=g^{\alpha\beta}g^{\gamma\delta}
(\partial_\alpha \partial_\gamma A)(\partial_\beta \partial_\delta
B)
\end{equation}
and is symmetric if one makes the changement $A \leftrightarrow B$.
The second generalized Poisson bracket acts on three phase space quantities
as
\begin{equation}
\{A;B,C\}=g^{\alpha\beta}g^{\gamma\delta}
(\partial_\alpha \partial_\gamma A)(\partial_\beta B)
(\partial_\delta C).
\end{equation}

\section{Gradient expansion of physical quantities}
\label{sec:gradexpofphysq}

In the previous section we have seen that the one particle 
Green's function $G$ can be written as a formal power series in $\hbar$,
where the correction terms $G_j$ in (\ref{eq:Gformexp}) 
are given by the partial
fraction series (\ref{eq:grab}). Performing the $\omega$ integal
in Eq. (\ref{eq:hcalfromG}) it is easy to see that only the
first order poles located on the upper half of the complex omega plane
give contributions to $h(\mathbf{R},\mathbf{p})$. Correspondingly,
by Eq. (\ref{eq:ncalc}) the density $n(\mathbf{R})$ has the expansion
\begin{equation}
n(\mathbf{R}) =
\sum_{j=0}^\infty\hbar^j
\int \!\!\frac{d^3p}{(2\pi\hbar)^3}B_{j,1}(\mathbf{R},\mathbf{p})
\equiv \sum_{j=0}^\infty\hbar^j g_j(\mathbf{R})
\label{eq:pertserofnr}
\end{equation}
Similarly, Eqs. (\ref{eq:Fformexp}), (\ref{eq:fcd}) and 
(\ref{eq:chicalfromF}) and (\ref{eq:gapcalcfromchi}) lead to
\begin{gather}
\Delta(\mathbf{R},\mathbf{p})\equiv \sum_{j=0}^\infty\hbar^j
f_j(\mathbf{R},\mathbf{p}) \nonumber\\
=
\sum_{j=0}^\infty\hbar^j \! \int d^3r e^{-i\mathbf{p}\mathbf{r}/\hbar}
v(\mathbf{r})\int \!\!\frac{d^3q}{(2\pi\hbar)^3} 
e^{-i\mathbf{q}\mathbf{r}/\hbar}
D_{j,1}(\mathbf{R},\mathbf{q}).
\label{eq:gaprpexp}
\end{gather}
Calculating the first few $g_j(\mathbf{R})$'s and 
$f_j(\mathbf{R},\mathbf{p})$'s it can be seen that the $\mathbf{R}$ 
dependence enters
in $g_j$ and $f_j$ through the quantities $U(\mathbf{R})$ and
$\Delta(\mathbf{R},\mathbf{p})$ and through the spatial derivatives
of order $\le j$ of $U(\mathbf{R})$ and $\Delta(\mathbf{R},\mathbf{p})$.
For $j=0$ there are no spatial derivatives (See Eqs. (\ref{eq:a01b01}) and
(\ref{eq:c01d01})). For $j=1$ the Poisson-brackets in (\ref{eq:a11b11})
bring the dependence also on gradients of $U(\mathbf{R})$ 
and $\Delta(\mathbf{R},\mathbf{p})$  
into $B_{1,1}$, and correspondingly into $g_1(\mathbf{R})$
for  complex $\Delta$. $B_{1,1}$ vanishes if $\Delta$ real, and
we consider in the following only this case.
Here we write the $j=2$ results expressed in terms of 
the generalized Poisson brackets
\begin{gather}
B_{2,1}(\mathbf{R},\mathbf{p})=
-\frac{2\nu^3\Delta-3\nu\Delta^3}{16E^7}\{\Delta;\nu,\nu\}
-\frac{3\Delta^2\nu}{32E^5}\{\nu,\nu\}_+
\nonumber\\
-\frac{2\nu\Delta^3-3\nu\Delta^2}{16E^7}\{\Delta;\Delta,\Delta\}
-\frac{3\nu^2\Delta^2-\nu^4-\Delta^4}{8E^7}\{\Delta;\nu,\Delta\}
\nonumber\\
-\frac{\Delta^4-4\nu^2\Delta^2}{16E^7}\{\nu;\nu,\nu\}
-\frac{\nu(2\nu^2-3\Delta^2)\Delta}{8E^7}\{\nu;\Delta,\nu\}
\nonumber\\
-\frac{5\nu^2\Delta^2}{16E^7}\{\nu;\Delta,\Delta\}
+\frac{(2\nu^2-\Delta^2)\Delta}{16E^5}\{\nu,\Delta\}_+
\nonumber\\
+\frac{\nu^3+\nu\Delta^2}{16E^7}\{\nu,\Delta\}^2
-\frac{\nu^3-2\nu\Delta^2}{32E^5}\{\Delta,\Delta\}_+,
\label{eq:g2polusok}
\end{gather} 
\begin{gather}
D_{2,1}(\mathbf{R},\mathbf{p})=
\frac{(3\Delta^3-2\nu^2)\nu\Delta}{16E^7} \{ \nu;\nu,\nu\}-
\frac{3\nu^2\Delta}{32E^5}\{\Delta,\Delta\}_+
\nonumber\\
-\frac{(\Delta^2-2\nu^2)\Delta}{32E^5}\{\nu,\nu\}_+
-\frac{\nu(2\Delta^2-3\nu^2)\Delta}{8E^7}\{\Delta;\Delta,\nu\}
\nonumber\\
-\frac{\nu(2\Delta^2-3\nu^2)\Delta}{16E^7}\{\nu;\Delta,\Delta\}-
\frac{\nu(\nu^2-2\Delta^2)}{16E^5}\{\nu,\Delta\}_+
\nonumber\\
-\frac{5\nu^2\Delta^2}{16E^7}\{\Delta;\nu,\nu\}+
\frac{\nu^4+\Delta^4-3\nu^2\Delta^2}{8E^7}\{\nu;\Delta,\nu\}
\nonumber\\
-\frac{\nu^4-4\nu^2\Delta^2}{16E^7}\{\Delta;\Delta,\Delta\}
+\frac{\Delta}{16E^5}\{\nu,\Delta\}^2.
\label{eq:f2polusok}
\end{gather}
For $\delta$ interaction Eq.(\ref{eq:gaprpexp}) simplyfies,
$f_j(\mathbf{p},\mathbf{R})$ has no momentum dependence:
\begin{gather}
 \Delta(\mathbf{R})\equiv \sum_{j=0}^\infty\hbar^j 
 f_j(\mathbf{R})\nonumber\\
=\frac{4\pi\hbar^2a}{m}\sum_{j=0}^\infty\hbar^j
  \int\frac{d^3 p}{(2\pi\hbar)^3}D^{reg}_{j,1}(\mathbf{R},\mathbf{p}),
\label{eq:pertserofmur}
\end{gather}
where $D^{reg}_{j,1}$ denotes the regularized part of $D_{j,1}$,
which can be obtained from Eq. (\ref{eq:gaprpexp}) if the pseudo-potential
is used for the interaction. 
Equations (\ref{eq:pertserofnr}) and (\ref{eq:pertserofmur}) can be
solved perturbatively whose formal {\sl solutions} become of the form
\begin{eqnarray}
n(\mathbf{R})&=& \sum_{j=0}^\infty \hbar^j n_j(\mathbf{R}) 
\label{eq:nsolser} \\
\Delta(\mathbf{R})&=&\sum_{j=0}^\infty \hbar^j \Delta_j(\mathbf{R})
\label{eq:deltsolser}
\end{eqnarray}

It is important to stress, that on the right hand sides of Eqs. 
(\ref{eq:pertserofnr}) and (\ref{eq:pertserofmur}) all the quantities
$g_j$ and $f_j$ depend on the total $\Delta(\mathbf{R})$
and $U(\mathbf{R})$, thus $n_j\ne g_j$ and $\Delta_j \ne f_j$.

\subsection{Local density approximation}

\begin{figure}
\centerline{
\epsfig{file=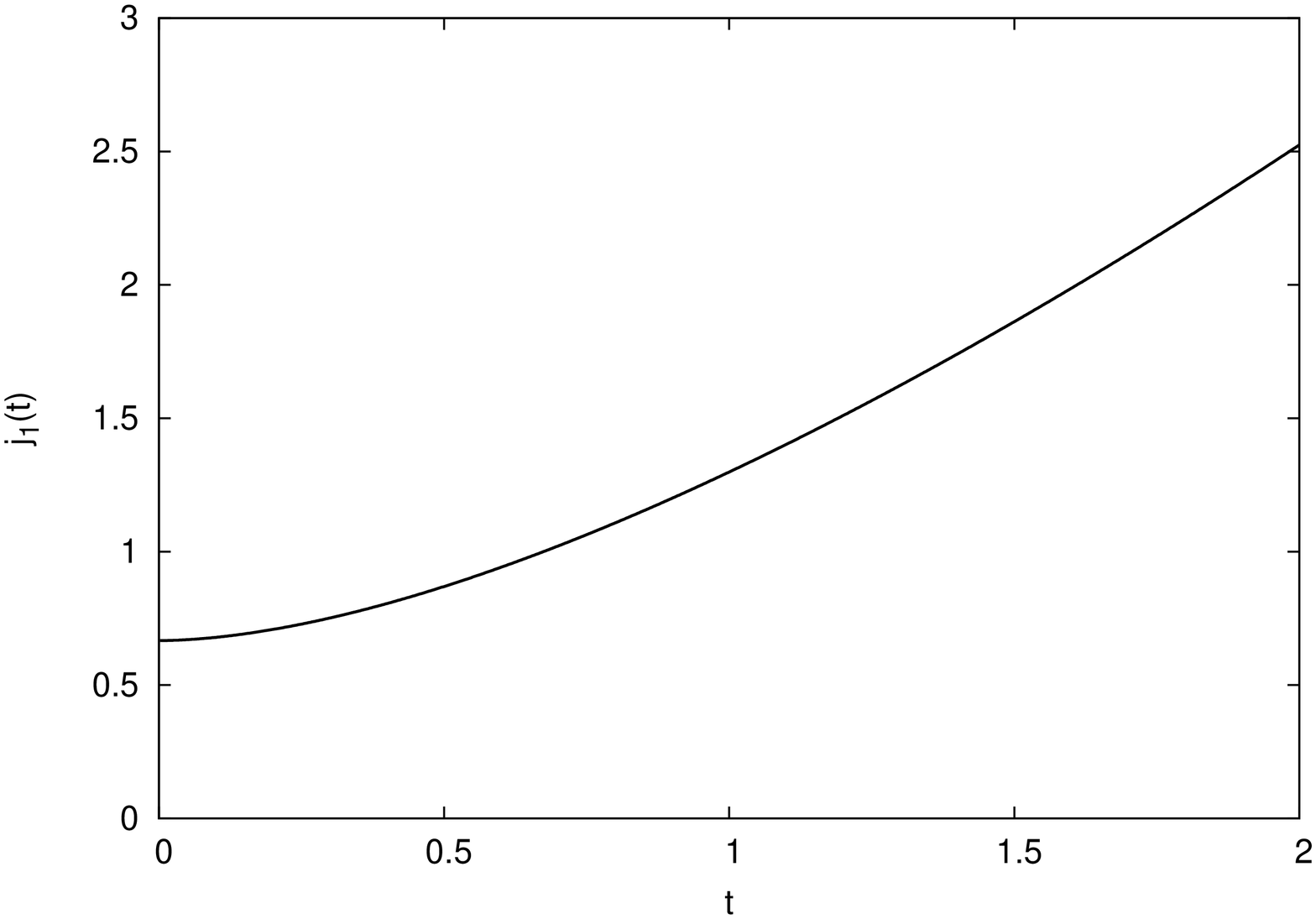,height=\hsize,angle=270}}
\caption{The dimensionless function $j_1(t)$ defined in
  Eq. (\ref{eq:j1}). \label{fig:j1t}} 
\end{figure}
The leading order $j=0$ approximation in Eqs. (\ref{eq:pertserofnr}) and
(\ref{eq:pertserofmur}) are equivalent to the LDA. In that approximation
one has to solve the equations
\begin{eqnarray}
n_0(\mathbf{R})&=&g_0(U_0(\mathbf{R}),\Delta_0(\mathbf{R})) 
\label{eq:zeroordg}\\
\Delta_0(\mathbf{R})&=&f_0(U_0(\mathbf{R}),\Delta_0(\mathbf{R})).
\label{eq:zeroordf}
\end{eqnarray}
The Hartree-Fock terms in $U(\mathbf{R})$ are density dependent. By
the notation $U_0$ in that case we mean that they are evaluated 
using $n_0(\mathbf{R})$.
If the Hartree-Fock terms are neglected $U_0=U_{ext}$ and $U_j=0$
for $j>0$.
Let us introduce the local chemical potential $\alpha$ by
\begin{equation}
\alpha(\mathbf{R})=\mu-U(\mathbf{R}).
\label{eq:alpha_def}
\end{equation}
The quantity $\nu(\mathbf{R},\mathbf{p})$ defined in Eq.(\ref{eq:nurplambda}) 
is simply $\nu(\mathbf{R},\mathbf{p})=p^2/2m-\alpha(\mathbf{R})$.
The phase space quantity $E(\mathbf{R},\mathbf{p})$ in 
Eq.(\ref{eq:ERpdef}) with real $\Delta(\mathbf{R})$ is equal to 
$E(\mathbf{R},\mathbf{p})=\sqrt{\nu^2(\mathbf{R},\mathbf{p})+
\Delta(\mathbf{R})^2}$. 
The $g_0$ function occuring in (\ref{eq:zeroordg}) can be calculted
from (\ref{eq:pertserofnr}) and (\ref{eq:a01b01}) and it is given by
\begin{equation}
 g_0(U(\mathbf{R}),\Delta(\mathbf{R}))=
\frac{1}{2}\int\frac{d^3 p}{(2\pi\hbar)^3}
  \left(1-\frac{\nu(\mathbf{R},\mathbf{p})}{E(\mathbf{R},\mathbf{p})}\right).
\label{eq:reszecske}  
\end{equation}
The momentum integrals can be performed analytically 
\begin{equation}\label{eq:suri2}
g_0(U(\mathbf{R}),\Delta(\mathbf{R}))=
\frac{1}{4{\pi}^2}\left(\frac{2m\alpha(\mathbf{R})}
{\hbar^2}\right)^{3/2}j_1\left(\frac{\Delta(\mathbf{R})}{\alpha(\mathbf{R})}
\right), 
\end{equation}
where we have used the dimensionless function $j_1(x)$ (See 
Appendix \ref{sec:integrals} and fig.\ref{fig:j1t}). 
The $f_0$ function in (\ref{eq:zeroordf}) can be calculated in a
similar way:
\begin{equation}\label{eq:divgap}
f_0(U(\mathbf{R}),\Delta(\mathbf{R}))=
-\frac{\Delta(\mathbf{R})}{2}g \!\! \int \!\!\frac{d^3 p}{(2\pi\hbar)^3}
\left[ \frac{1}{E(\mathbf{R},\mathbf{p})}-
\frac{2m}{p^2} \right].
\end{equation}
The second term in the integrand ensures a finite value for the momentum
integral, i.e., $D_{0,1}$ is regularized with this term substracted.
The momentum integral in Eq. (\ref{eq:divgap})
can be written in terms of complete elliptic functions
(see Appendix~\ref{sec:integrals}) and can be expressed 
for negative scattering 
length as  
\begin{equation}
f_0(U(\mathbf{R}),\Delta(\mathbf{R}))
=\Delta(\mathbf{R})
\frac{2}{\pi}|a|\sqrt{\frac{2m\alpha(\mathbf{R})}{\hbar^2}} \, i_1\!
\left(\frac{\Delta(\mathbf{R})}{\alpha(\mathbf{R})}\right)
\label{eq:gapi1},
\end{equation}
\begin{figure}
\centerline{
\epsfig{file=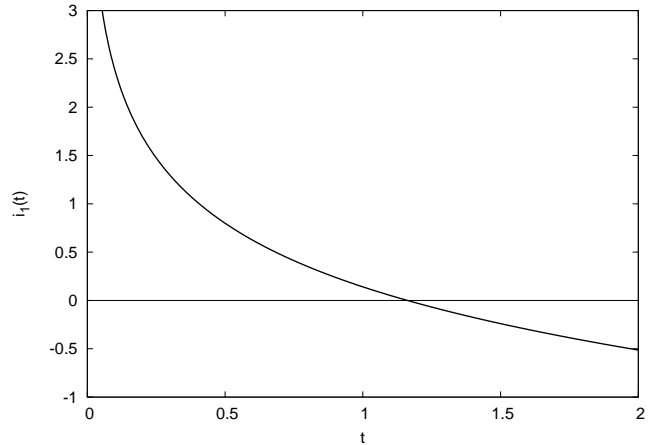,height=\hsize,angle=270}}
\caption{The dimensionless function $i_1(t)$ defined in
  Eq. (\ref{eq:i1}).}
\label{fig:i1t} 
\end{figure}
\noindent where the dimensionless function $i_1(x)$, depicted on
fig.~\ref{fig:i1t},  is defined by Eq. (\ref{eq:i1}). 
The overall constant chemical potential $\mu$ is fixed by
\begin{equation}
\frac{N}{2}=\int n_0(\mathbf{R})d^3 R.
\label{eq:norm0}
\end{equation}
Here $N$ is the total particle number (including both hyperfine states). 
Solutions to Eqs. (\ref{eq:zeroordg}), (\ref{eq:zeroordf}) and (\ref{eq:norm0})
with $g_0$ and $f_0$ given by (\ref{eq:suri2}) and (\ref{eq:gapi1})
are the solutions in leading order.
Thus, the zeroth order terms in the gradient expansion lead to the Local
Density Approximation. This corresponds to he Thomas-Fermi approach
generalized to taking into account the pairing field
$\Delta(\mathbf{r})$.
In the following we 
shall calculate corrections to $g_0$ and $f_0$. 

\subsection{$O(\hbar)$ order}

We have shown that for real $\Delta$ the 
quantity $B_{1,1}$ is zero. Thus, $g_1(\mathbf{R})=0$ in 
Eq. (\ref{eq:pertserofnr}). Due to the property (\ref{eq:c11d11}) the 
regularized part of $D_{1,1}$ is also zero. It means there are no corrections 
to the density and to the gap equations in this $j=1$ order.

\subsection{$O(\hbar^2)$ order}

The evaluations of the $j=2$ second order corrections 
$g_2(\mathbf{R})$, $f_2(\mathbf{R})$ are rather tedious. In case of 
momentum independent gap and $\nu(\mathbf{R},\mathbf{p})$ given by 
Eq. (\ref{eq:nurplambda})
nonvanishing generalized Poisson brackets are
\begin{equation}
 \{\nu,\Delta\}^2=\sum_{i,j=1}^3\frac{p_i}{m}\frac{p_j}{m} 
\left(\nabla\Delta\right)_i\left(\nabla\Delta\right)_j,
\end{equation}
\begin{equation}
  \{\nu,\nu\}_+=\frac{2}{m}\left(\nabla^2 U\right), \quad 
\{\nu,\Delta\}_+=\frac{1}{m}\left(\nabla^2\Delta\right),
\end{equation}
\begin{eqnarray}
\{\nu;\nu,\nu\}&=&\frac{1}{m}\left(\nabla U\right)^2+\sum_{i,j=1}^3
\left[\left(\frac{\partial^2 U}{\partial R_i\partial R_j}\right)
\frac{p_i}{m}\frac{p_j}{m}\right], \nonumber\\ 
\{\nu;\Delta,\nu\}&=&\frac{1}{m}\left(\nabla \Delta\right)
\left(\nabla U\right),
\nonumber\\ 
\{\Delta;\nu,\nu\}&=&\frac{1}{m^2}\sum_{i,j=1}^3
\left(\frac{\partial^2\Delta}{\partial R_i\partial R_j}
\right)p_ip_j,\nonumber\\  
\{\nu;\Delta,\Delta\}&=&\frac{1}{m}\left(\nabla\Delta\right)^2.
\label{eq:poissonbra}
\end{eqnarray}
Note that the Laplace operator will be written as $\nabla^2$ to avoid
confusion with the gap.
We gave a general expression for $B_{2,1}$ in Eq. (\ref{eq:g2polusok}) by 
which $g_2$ can be obtained by evaluating the momentum integrals. For the 
details see Appendix \ref{sec:integrals}. The result is rather lenghty and 
can be presented in the following way. Let us define
\begin{equation}
A(\alpha) \equiv A(\alpha(\mathbf{R}))=\frac{(2m\alpha(\mathbf{R}))^{3/2}}
{2\pi^2\hbar^3},
\label{eq:aalpha}
\end{equation}  
and the dimensionless combination $t$ by
\begin{equation}
t \equiv t(\mathbf{R})=\frac{\Delta(\mathbf{R})}{\alpha(\mathbf{R})},
\label{eq:tdef}
\end{equation} 
where $\alpha(\mathbf{R})$ is the local chemical potential 
(\ref{eq:alpha_def}).
The second order correction $g_2$ to the density equation 
(\ref{eq:pertserofnr}) is
\begin{gather}
g_2(\mathbf{R}) = A(\alpha)\left[\frac{(\nabla^2 U)}{m\alpha^2}H_1(t)
+\frac{(\nabla\Delta)(\nabla U)}{m\alpha^3}H_2(t)\right. \nonumber\\
\left.+\frac{(\nabla\Delta)^2}{m\alpha^3}H_3(t) 
+\frac{(\nabla^2\Delta)}{m\alpha^2}H_4(t)
+ \frac{(\nabla U)^2}{m\alpha^3}H_5(t)\right]. 
\label{eq:n2h}
\end{gather}
Functions $H_1(t), \ldots,H_5(t)$ are given in Appendix \ref{sec:functions} 
by Eq. (\ref{eq:hak}).
The second order corrections to the gap equation can be calculated using 
Eqs. (\ref{eq:f2polusok}) and (\ref{eq:pertserofmur}). In second order the 
momentum integrals exsist, consequently $D_{2,1}^{reg}=D_{2,1}$, i.e., there
is no need to regularize $D_{2,1}$. Proceeding as above, the momentum 
integrals can be treated as in Appendix \ref{sec:integrals}.  
Second order gradient correction $f_2$ to the gap equation 
(\ref{eq:pertserofmur}) can be expressed as 
\begin{gather}
f_2(\mathbf{R})=g A(\alpha)\left[\frac{(\nabla^2 U)}{m \alpha^2}M_1(t)
+\frac{(\nabla\Delta)(\nabla U)}{m \alpha^3}M_2(t)\right.\nonumber\\
\left.+\frac{(\nabla\Delta)^2}{m \alpha^3}M_3(t) \!
+\frac{(\nabla^2\Delta)}{m \alpha^2}M_4(t)+\!
\frac{(\nabla U)^2}{m \alpha^3}M_5(t)\right],
\label{eq:mu2M}
\end{gather}
where the functions $M_1(t), \ldots, M_5(t)$ also can be found in Appendix 
\ref{sec:functions} in Eq. (\ref{eq:mek}).
From Eqs. (\ref{eq:n2h}) and (\ref{eq:mu2M}) it is obvious that the second 
order corrections to the density and to the gap equation involve the 
spatial derivatives of the external potential $U_{ext}$ and the gap 
profile $\Delta$. If the generalized Hartree-Fock approximation is
considered $g_2$ and $f_2$ will contain terms with the spatial derivatives
of the density, too. 

In second order approximation $n\approx n_0+\hbar^2n_2$, $\Delta\approx
\Delta_0+\hbar^2\Delta_2$, $U\approx (U_{ext}+gn_0)+\hbar^2 gn_2$. Expanding
both sides of Eqs. (\ref{eq:pertserofnr}) and (\ref{eq:pertserofmur})
up to second order in $\hbar$, the zeroth order terms cancel. The
second order gradient corrections to density and the gap are the
solution of the 
\begin{equation}
\left(\begin{array}{cc}
\! \!1-\partial_1 g_0(U_0,\Delta_0) & -\partial_2 g_0(U_0,\Delta_0)\\
-\partial_1 f_0(U_0,\Delta_0) & 1-\partial_2 f_0(U_0,\Delta_0)\!\!
\end{array}\right)\!\!
\left(\begin{array}{c}
n_2\! \\ \!\Delta_2 \!
\end{array}\right)=
\left(\begin{array}{c}
\! g_2 \! \\ \!f_2 \!
\end{array}\right)
\label{eq:secondordercalc}
\end{equation}
inhomogeneous linear equations (Here $\partial_1$ and $\partial_2$
denote partial derivatives with respect to $n_0$ and $\Delta_0$, respectively).

The spatial derivatives of the 
density in Eqs. (\ref{eq:n2h}) and (\ref{eq:mu2M}) 
are however missing if the MF-BCS model is considered, which
neglects the Hartree-Fock terms in $\nu$.
The density and the gap still get gradient corrections in this model,
which case will be studied next.

\section{Second order corrections calculated perturbatively
         in the MF-BCS model}
\label{sec:secordercorr}

In MF-BCS model the quantity 
\begin{equation}
\alpha(\mathbf{R})=\mu-U_{ext}(\mathbf{R})
\end{equation}
is density independent and it is advantageous to use $t$ defined
in (\ref{eq:tdef}) instead of $\Delta$. We keep the density equation
(\ref{eq:pertserofnr}), but rewrite (\ref{eq:pertserofmur}) as
\begin{equation}
t(\mathbf{R})=\sum_{j=0}^\infty \tilde{f}_j \hbar^j,\quad 
\tilde{f}_j=f_j/\alpha
\label{eq:leggapeq}
\end{equation}
$\tilde{f}_1$ vanishes as in the previously
discussed general case. Similarly to (\ref{eq:deltsolser})
we are seeking the solution for $t(\mathbf{R})$ as a formal
series in $\hbar$
\begin{equation}
t(\mathbf{R})=\sum_{j=0}^\infty \hbar^j t_j(\mathbf{R}),
\label{eq:tsolser}
\end{equation}
where $t_1$ is zero due $\tilde{f}_1=0$.
The first $\tilde{f}_0$ is given by
\begin{equation}
\tilde{f}_0=\frac{2}{\pi}|a|
\left(\frac{2m \alpha(\mathbf{R})}{\hbar^2}\right)^{1/2} t i_1(t),
\end{equation}
and $\tilde{f}_2$ can be obtained from Eq. (\ref{eq:mu2M}) by dividing
both sides by $\alpha$. $g_0$ is still given by (\ref{eq:suri2})

In the MF-BCS model the leading order LDA equations are
\begin{eqnarray}
n_0(\mathbf{R})&=& \frac{1}{4\pi^2}
\left(\frac{2m \alpha(\mathbf{R})}{\hbar^2}\right)^{3/2}j_1(t_0), \\
1&=&\frac{2}{\pi}|a|
\left(\frac{2m \alpha(\mathbf{R})}{\hbar^2}\right)^{1/2}i_1(t_0).
\label{eq:gapeqleg}
\end{eqnarray}
For fixed chemical potential $\mu$ the $t_0(\mathbf{R})$ profile
can be calculated from (\ref{eq:gapeqleg}).

   
To obtain the second
order gradient corrections $n_2$ and $t_2$ we can approximate
in the expressions (\ref{eq:n2h}), (\ref{eq:mu2M}) for $g_2$ and $\tilde{f}_2$
the quantity $t$ by its zeroth order value $t_0$, because
$g_2$ and $\tilde{f}_2$ are already of second order.
Taking the gradient of the leading order gap equation (\ref{eq:gapeqleg}) 
$\nabla \Delta(\mathbf{R})$ can be approximated as
\begin{equation}
(\nabla\Delta_0(\mathbf{R}))=-(\nabla U_{ext}(\mathbf{R}))T_1(t_0), 
\label{eq:gradmugradU}
\end{equation}
where the dimensionless function $T_1(t)$ is given by 
\begin{equation}
T_1(t)=\frac{j_3(t)}{t\, i_3(t)}.
\end{equation} 
(See Appendix \ref{sec:integrals} for the definitions of $i_n(t)$ and 
$j_n(t)$). Taking the divergence of the two 
sides of Eq. (\ref{eq:gradmugradU})
$\nabla^2 \Delta_0(\mathbf{R})$ can be reduced to
\begin{equation}
\nabla^2\Delta_0(\mathbf{R})=-(\nabla^2 U_{ext})T_1(t_0)+\frac{(\nabla
    U_{ext})^2}{\alpha}T_2(t_0),
\label{eq:lapmufromu}
\end{equation}
where we have introduced an other dimensionless $t$-dependent function 
$T_2(t)$. Explicitely:
\begin{eqnarray}
T_2(t)&=&\left[t-T_1(t)
\right]\nonumber \\
&&\times\frac{i_3(t)j_3(t)+3t^2\left(j_5(t)i_3(t)-j_3(t)i_5(t)\right)}{t^2\cdot
    i^2_3(t)}.\nonumber \\
\end{eqnarray}
Using Eqs. (\ref{eq:gradmugradU}) and (\ref{eq:lapmufromu}) in the expressions 
of the second order corrections (\ref{eq:n2h}) and (\ref{eq:mu2M}) and the 
results of Appendix \ref{sec:integrals} and \ref{sec:functions} the 
perturbatively calculated corrections are 
\begin{equation}\label{eq:b2font}
g_2(\mathbf{R})=\frac{A(\alpha)}{m\alpha^2} \left[
\frac{(\nabla U_{ext})^2}{\alpha}P_1(t_0)+(\nabla^2 U_{ext})Q_1(t_0) \right]
\end{equation}
and
\begin{equation}\label{eq:d2font}
f_2(\mathbf{R})=\frac{g A(\alpha)}{m\alpha^2} \left[
\frac{(\nabla U_{ext})^2}{\alpha}P_2(t_0)+(\nabla^2 U_{ext})Q_2(t_0) \right],
\end{equation}
where $A(\alpha)$ is defined by (\ref{eq:aalpha}).
Explicit expressions for $P_1$, $Q_1$, $P_2$ and $Q_2$ are given in 
Appendix \ref{sec:functions} in 
Eqs. (\ref{eq:p1})-(\ref{eq:q2}).
Corrections (\ref{eq:b2font}), (\ref{eq:d2font}) involve terms proportional to 
$(\nabla U_{ext})^2$ and $\nabla^2 U_{ext}$. We remind the reader
that $g$ is proportional to the scattering length $a$ (see Eq. 
(\ref{eq:contactinteraction})).

Expanding both sides of Eq. (\ref{eq:leggapeq}) up to second order
using $\tilde{f}_0(t)\approx \tilde{f}_0(t_0+\hbar^2 t_2)
\approx \tilde{f}_0(t_0)+
\hbar^2 t_2 \tilde{f}'_0(t_0)$
the zeroth order terms cancel, and $t_2$ can be expressed as
\begin{equation}
t_2=-\frac{\tilde{f}_2}{t_0 i_1'(t_0)}
 \frac{1}{\frac{2|a|}{\pi}
\left(\frac{2m \alpha(\mathbf{R})}{\hbar^2}\right)^{1/2}}
\label{eq:t2expr}
\end{equation}
Note that $t_2$ depends on the scattering length only through $t_0$.
The second order gradient correction of the gap is 
$\Delta_2(\mathbf{R})=t_2(\mathbf{R})\alpha(\mathbf{R})$ in the
MF-BCS model. 
Using the same approximation for the density the second
order gradient correction of the density is 
\begin{equation}
n_2=\frac{1}{4\pi^2}
\left(\frac{2m \alpha(\mathbf{R})}{\hbar^2}\right)^{3/2}t_2 j_1'(t_0)+g_2
\label{eq:n2expr}
\end{equation}
These are the first nontrivial gradient 
expansion terms.
The simplification in the MF-BCS model has arised from the fact that
in Eq. (\ref{eq:secondordercalc}) the $\partial_1$ derivatives (i.e.,
the derivatives with respect to the density) are zero.

\section{Universal prefactor of the von Weizs\"acker type correction}
\label{sec:univprefactor}

The present paper supply the derivation of some of the relations used
already in our earlier paper \cite{csordas07}. To compare with the
results of \cite{csordas07} one has to apply the limit $a \to \infty$
to Eqs. (\ref{eq:b2font}-\ref{eq:n2expr}) (see also Appendix B in
applying this limit). The leading order $t_0(\mathbf{R})$ profile is 
constant, which can be seen from Eq. (\ref{eq:gapeqleg}). Let us 
denote by T the root of $i_1(T)=0$,  
then 
\begin{equation}
t_0(\mathbf{R})=T \approx 1.1622, \quad \quad j_1(T)\approx 1.4688
\end{equation}
($T$ is defined by the requirement $i_1(T)=0$ in order that Eq.
(\ref{eq:gapeqleg}) remains meaningful in the limit $a \to
\infty$. See also (\ref{eq:specp1})-(\ref{eq:specq2})
. )
It leads to
\begin{eqnarray} 
2n(\mathbf{R})&=&(\mu-U_{ext})^{3/2}\frac{1}{2\pi^2}j_1(T)
\left(\frac{2m}{\hbar^2} \right)^{3/2}  \nonumber \\
&&\left(1 -\frac{\hbar^2}{2m}j_1(0)\frac{21}{48}\frac{\nabla^2
  U_{ext}}{(\mu-U_{ext})^{2}} \right.\nonumber\\
&&\left. -\frac{\hbar^2}{2m}j_1(0)\frac{21}{192}\frac{(\nabla
  U_{ext})^2}{(\mu-U_{ext})^{3}}\right). \label{eq:densforus}
\end{eqnarray}
Note that $n=n_\uparrow=n_\downarrow$ in our notation. 
Similarly from Eqs. (\ref{eq:tdef}), (\ref{eq:leggapeq}),
(\ref{eq:d2font}), (\ref{eq:t2expr}) one gets
\begin{eqnarray}
\Delta(\mathbf{R})&=&T(\mu-U_{ext})-\frac{4+7T^2}{36T}
\frac{\hbar^2}{2m}\frac{\nabla^2   
  U_{ext}}{(\mu-U_{ext})} \nonumber\\
&&-\frac{7T^2-8}{144T}\frac{\hbar^2}{2m}\frac{(\nabla
  U_{ext})^2}{(\mu-U_{ext})^{2}},  
\end{eqnarray}
Eq. (\ref{eq:densforus}) can be rewritten as
\begin{eqnarray}
\mu-U_{ext}&=&(j_1(T))^{-2/3}(2\pi^2)^{2/3}\frac{\hbar^2}{2m}(2n)^{2/3}
\nonumber\\
&&+\frac{7\hbar^2}{72m}\left\{
\frac{(\nabla (\mu -U_{ext}))^2}{4(\mu-U_{ext})^{2}}-\frac{\nabla^2
  (\mu-U_{ext})}{(\mu-U_{ext})}
\right\},\nonumber \\
\end{eqnarray}
where we have taken into account that the seecond term in the right
hand side of the equation is a correction. The first step of the
iteration on the right hand side leads after a rearrengement to the
Thomas-Fermi-Weizs\"acker type equation:
\begin{equation}
-\kappa_W\frac{\hbar^2}{2m}\frac{\nabla^2 n^{1/2}}{n^{1/2}}+\xi
\kappa_F (2n)^{2/3}+U_{ext}=\mu
\label{eq:weiztype} 
\end{equation}
Here
\begin{equation}
\kappa_W=\frac{7}{27}, \quad \xi=\left(\frac{2}{3j_1(T)}
\right)^{2/3}, \quad \kappa_F=\frac{\hbar^2}{2m}(3\pi^2)^{2/3}.
\end{equation}
Note that $\xi$ is the usual universal constant introduced for the
homogeneous system by the definition $\mu=\xi
\epsilon_F$. ($\epsilon_F$ being the Fermi energy $\hbar^2k_F^2/(2m)$,
where $k_F=(6\pi^2 n)^{1/3}$). The above value is valid in the MF-BCS 
model \cite{csordas07,Giorgini08}. Numerically $\xi=0.59$ in this model, while
the Monte Carlo Simulations have provided $\xi=0.37-0.44$
\cite{Bulgac10,Giorgini08}. Note, that for a normal system at
unitarity $\xi=0.55$ \cite{Carlson03,Carlson03a,Bulgac07}, the
corresponding MF-BCS value is $\xi=1$, i.e. free gas value, since
$\xi_N=\alpha+\beta$, where $\alpha$ is the ratio of the mass and the
effective mass (beeing unity in the MF-BCS model) and $\beta$ is zero
(see Ref. \cite{Bulgac07} and references therein).  
The first term on the left hand side of
Eq. (\ref{eq:weiztype}) is of the form of the von Weizs\"acker correction
to the Thomas-Fermi theory (see for the early history of the problem
Ref. \cite{Gombas49}). By now it is well established that 
$\kappa_W$=1
(originally derived by von Weizs\"acker) is the correct value in case of a
rapidly varying density with a small amplitude, while in case of a
smooth external potential $\kappa_W=1/9$. This value of $\kappa_W$ was
first derived by Kirzhnits \cite{Kirzhnits57} and by Kompaneets and
Pavlovskii \cite{Kompaneets57}. (See for reviews of the density
gradient expansions \cite{Parr89,Dreizler93,Brack97}). 

It is worth mentioning that $\kappa_W=1/9$ was found \cite{Meyer76}
the optimal value when the energy of a free gas in a harmonic
oscillator potential was compared with the quantum mechanical result
via second order perturbation theory. This suggests that such an
external potential occuring in trapped gases is well suited for a
gradient expansion of the density.

There has been a
renewed interest in recent years concerning the von Weizs\"acker
correction in case of the trapped unitary Fermi gas
\cite{Kim04,Manini05,Salasnich08,Salasnich08a,Adhikari08,Salasnich09,Wen08,Adhikari09,Rupak09,Zubarev09}. The value of
$\kappa_W=1$ has been chosen in \cite{Kim04,Manini05}, while
$\kappa_W=1/4$ has been obtained in
\cite{Salasnich08,Salasnich08a,Salasnich09,Wen08} by assuming the
validity of a kind of Ginzburg-Landau theory at zero
temperature. Furthermore, an expansion in powers of $d=4-\epsilon$
spatial dimensions has led to $\kappa_W=0.176$ \cite{Rupak09} by
extrapolating the result to three dimensions. A comparison between the
choices $\kappa=1/9$ and $\kappa=1/4$ has been carried out in
\cite{Zubarev09} by studying fermion systems at unitarity with
particle numbers up to 50. It has been found that the choice
$\kappa_W=1/4$ provides better results for the energy except at few
particle numbers. This finding backs our result for $\kappa_W=7/27$,
which is quite close to this value.

\section{Isotropic harmonic trapping} 
\label{sec:isotropictrap}

\begin{figure}
\centerline{
\epsfig{file=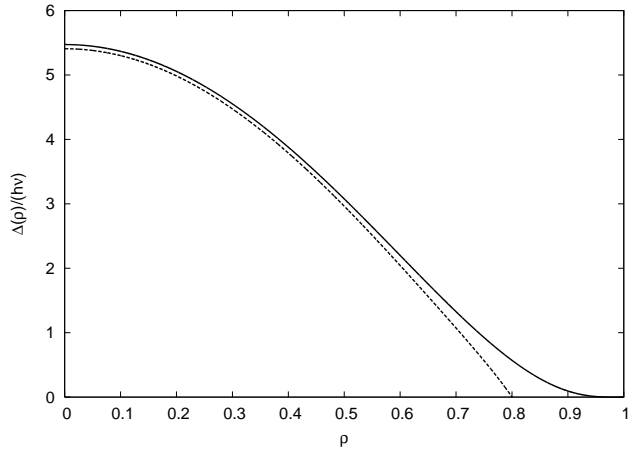,height=\hsize,angle=270}
}
\caption{Solutions of the gap equation without (solid line) and with 
(dashed line) the second order gradient corrections as a function of the
dimensionless radius $\rho=R/R_{TF}$, and measured in units of
$\hbar\omega_0=h\nu$ (see text). Parameters are: $a/d=1/(3\pi)$, $R_{TF}/d=8$.
\label{fig:gap}}
\end{figure}

\begin{figure}
\centerline{
\epsfig{file=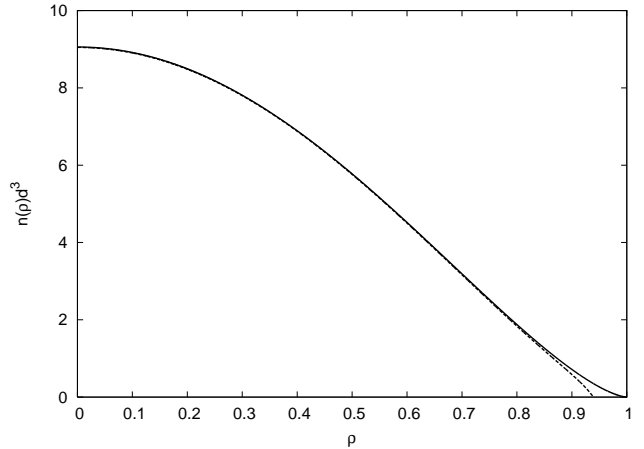,height=\hsize,angle=270}
}
\caption{Density profiles without (solid line) and with 
(dashed line) the second order gradient corrections as a function of the
dimensionless radius $\rho=R/R_{TF}$, and measured in units of $1/d^3$
Parameters are: $a/d=1/(3\pi)$, $R_{TF}/d=8$.
\label{fig:sur}}
\end{figure}

As an application of
  Eqs. (\ref{eq:b2font})-(\ref{eq:n2expr})
let us apply our results to the special case of 
isotropic harmonic trapping potential 
\begin{equation}
U(\mathbf{R})=\frac{1}{2}m \omega^2_0 R^2.
\label{eq:expot}
\end{equation}
In local density approximation the Thomas Fermi radius $R_{TF}$ is 
introduced by the relation 
\begin{equation}
\mu=\frac{1}{2}m \omega^2_0 R_{TF}^2,
\end{equation}  
which ensures $n_0(R_{TF})=0$.
It is advantageuos to use the dimensionless combination  
\begin{equation}
\varrho=R/R_{TF}
\end{equation}
for the radial distance.
A natural characteristic length of the harmonic oscillator problem is the
oscillator length $d=\sqrt{\hbar/(m\omega_0)}$. 
The LDA gap equation (\ref{eq:gapeqleg}) for harmonic confinement  
\begin{equation}
1=\frac{2}{\pi}\frac{|a|R_{TF}}{d^2}(1-\varrho^2)^{1/2}i_1(t_0)
\label{eq:gapharleg}
\end{equation}
provides us a profile $t_0(\varrho)$, which depends on the single dimensionless
parameter $|a|R_{TF}/d^2$.
Up to second order gradient
corrections the gap can be expressed as
\begin{equation}
\frac{\Delta(\varrho)}{\hbar\omega_0}\cong\frac{1}{2}
\left( \frac{R_{TF}}{d}\right)^2(1-\varrho^2)\left(t_0(\varrho)+
  \frac{d^4}{R_{TF}^4} \delta \tilde{t}(\varrho)\right),
\end{equation}
where $\delta \tilde{t}$ can be read off from (\ref{eq:t2expr}) as
\begin{equation}
\delta t = \frac{1}{t_0 i_1'(t_0)}
\left(\frac{16 \varrho^2 P_2(t_0)}{(1-\varrho^2)^3}+
\frac{24 Q_2(t_0)}{(1-\varrho^2)^2} \right).
\end{equation}
Eq. (\ref{eq:n2expr}) together with the leading LDA for the
density can be expressed as
\begin{equation}
d^3 n(\varrho)\cong\frac{(1-\varrho^2)^{3/2}}{4\pi^2}
\frac{R_{TF}^3}{d^3} 
\left[j_1(t_0)+\frac{d^4}{R_{TF}^4}
\delta \tilde{n}(\varrho,t_0) \right],
\end{equation}
where
\begin{eqnarray}
\delta \tilde{n}(\varrho,t_0)&=&
\frac{16 \varrho^2}{(1-\varrho^2)^3}
\left(P_1(t_0)+\frac{j_1'(t_0)}{t_0 i_1'(t_0)} P_2(t_0)\right)\nonumber\\
&+&\frac{24 }{(1-\varrho^2)^2}
\left(Q_1(t_0)+\frac{j_1'(t_0)}{t_0 i_1'(t_0)} Q_2(t_0)\right)\!\! .
\end{eqnarray}

It is clearly seen that the small 
parameter of the problem is $d/R_{TF}$. The magnitude of the correction as 
compared to the leading term is 
proportional to $(d/R_{TF})^4$ both for the density and the gap.  




At the Feshbach point ($a \to \infty$) our results can be 
further simplify 

\begin{equation}
\delta \tilde{t}=-\frac{7T^2-8}{36T}\frac{\varrho^2}{(1-\varrho^2)^3}
-\frac{7T^2+4}{6T}\frac{1}{(1-\varrho^2)^2},
\end{equation}
\begin{equation}
\delta \tilde{n}= -7 j_1(T)\left(
  \frac{1}{24}\frac{\varrho^2}{(1-\varrho^2)^3} +
  \frac{1}{4}\frac{1}{(1-\varrho^2)^2}\right). 
\end{equation}

$\delta \tilde{t}$ and $\delta \tilde{n}$ are universal at unitarity
(at the Feshbach resonance) for a spherical parabolic trap: they do not contain
any parameter of the 
two particle interaction. 

\section{Summary and Conclusions}
\label{sec:summary}

We have calculated the gradient corrections on the BCS side of the
Feshbach resonance to the generalized Thomas-Fermi model, which
represents the LDA in the presence of pairing. Though the correction
terms have a prefactor, which is small for typical trap potentials
already at moderately large particle numbers, the corrections get
large due to the singularities at the LDA border of the cloud. At
unitarity 
a von Weizs\"acker type correction appears whose universal prefactor has
been derived as $\kappa_W=7/27$. This value is quite close to $1/4$
proposed in refs. \cite{Salasnich08,Salasnich09,Wen08,Adhikari09} and
is also not far from the $\epsilon$-expansion result \cite{Rupak09} as
extrapolated to three dimensions.

It is remarkable that by inverting the functional $n[U_{ext}]$ to
order $\hbar^2$, as it has been done in Sec. \ref{sec:univprefactor},
the singularities at $\mu=U_{ext}$ disappear and the density can be
continued to infinity. This situation is similar to what happens in
case of the free gas, and perhaps, the most physical justification is,
which starts the calculation at finite temperature and the zero
temperature limit is taken at the end  \cite{Dreizler93,Perrot79,Bartel85}. One has to keep in
mind, however, that it does not mean that even the asymptotic decay of
the density follow the true one in general.  

Away from unitarity, however, the situation is much more complicated
and needs further study. Instead then one can use the treatment
applied in Sec. \ref{sec:isotropictrap} (i.e., to regard the gradient
terms as corrections and keep away from the surface region, which
becomes, however, larger and larger when tending to the BCS limit).

In Figure~\ref{fig:gap} we have depicted the gap profile
$\Delta(\varrho)/\hbar\omega_0$ both in LDA and with gradient
corrections. At a certain radius $\varrho_2$ the gap with gradient
corrections becomes zero. 

In Figure~\ref{fig:sur} we show the dimensionless density profile
at the same parameters as for Figure~\ref{fig:gap}. 
The deviation from the LDA profile is much less pronounced 
at those particular parameters  in the
region where the gradient expansion is applicable. Note that in the
figures both curves are calculated at the same $\mu$ values, so they
belong to slightly different particle numbers.

The distance $\varrho_1$ from the origin, where
$\Delta(\varrho_1)/\hbar\omega_0=1$, decreases when the magnitude of
the scattering length becomes shorter, which means that the most
suitable situation exists at the Feshbach resonance. 
In the weak coupling (BCS) limit $\Delta$ is smaller
than $\hbar\omega_0$ already at the point $\mathbf{r}=0.$ For
$\varrho>\varrho_1$ the $\Delta(\varrho)/\hbar\omega_0$ function
steeply goes to zero (see Fig.~\ref{fig:gap}) beyond which point even
its formal continuation becomes meaningless reflecting 
the fact that such an expansion is not adequate when the gap function
$\Delta(\mathbf{r})$ gets smaller than the level spacing of the trap. 
One has to emphasize that
this behavior has been shown when the first nonzero correction is
treated perturbatively. More generally, the 
solution levels off for increasing $\varrho$ and one can define the
radius $\varrho_2$ in such a way $\Delta(\varrho)/\hbar\omega_0<
\delta$ for $\varrho>\varrho_2$ with $\delta$ as a suitable chosen
small parameter. Actually, in the region $\varrho>\varrho_2$ one has
to apply another method instead the one developed in this paper to get
more accurate result, but the difference might be small. This problem
goes beyond the scope of the present paper and planned as a
forthcoming work.  
 
{\it Note added:} After submission of the paper we have learned that
the density matrix in case of the inhomogeneous superfluid Fermi
systems was derived in Ref. \cite{Taruishi92} to
$o(\hbar^2)$ using the
Wigner-Kirkwood $\hbar$-expansion method by regarding the
pairpotential as an external one, which is an intermediate step in our
work (see also \cite{Szepfalusy64}). We are greatful to prof. Schuck
for informing us of the papers \cite{Groenewold46,Taruishi92}.

\begin{acknowledgments}
The present work has been partially supported by the Hungarian Scientific
Research Fund under Grant Nos. OTKA 77534/77629 and OTKA 75529.
\end{acknowledgments}

\appendix
\section{Momentum integrals}
\label{sec:integrals}

In the zeroth order two types of momentum integrals occur:
\begin{equation}
K_1=\int\frac{d^3 p}{(2\pi\hbar)^3}\left(\frac{1}{E}-
\frac{2m}{p^2}\right)=\frac{A(\alpha)}{\alpha}i_1(t)|_{t=\Delta/\alpha},
\label{eq:K1def}
\end{equation}
\begin{equation}
L_1=\int\frac{d^3 p}{(2\pi\hbar)^3}\left(1-\frac{\nu}{E}\right)=
A(\alpha)j_1(t)|_{t=\Delta/\alpha},
\label{eq:L1def}
\end{equation}
where $A(\alpha)$ has been introduced in (\ref{eq:aalpha}) with
$\alpha$ the local chemical potential (\ref{eq:alpha_def}).
The dimensionless integrals $i_1(t)$ and $j_1(t)$ are defined as
\begin{eqnarray}
 i_1(t)&=&\int_0^\infty d\, x
 \left(\frac{x^2}{\sqrt{(x^2-1)^2+t^2}}-1\right).\label{eq:i1}\\
j_1(t)&=&\int^\infty_0 x^2 d x \left(1-\frac{x^2-1}{\sqrt{(x^2-1)^2+t^2}}.
\right)\label{eq:j1}
\end{eqnarray}
$i_1(t)$ and $j_1(t)$ can be expressed in terms of complete elliptic 
integrals $K(k)$ and $E(k)$ (see Reference \cite{GradsteinRyzhick})
\begin{eqnarray}
K(k)&=&\int_0^{\pi/2} d \varphi \frac{1}{\sqrt{1-k^2\sin^2\varphi}}, \\
E(k)&=&\int_0^{\pi/2} d\varphi \sqrt{1-k^2\sin^2\varphi}, 
\end{eqnarray}
as
\begin{equation}
i_1(t)=\sqrt[4]{1+t^2}\left[K(k)-2E(k)\right]
\end{equation}
\begin{equation}
j_1(t)=\frac{1}{3}\sqrt[4]{1+t^2}\left[\frac{{t^2}K(k)}
{1+\sqrt{(1+t^2)}}+2E(k)\right],
\end{equation}
where the modulus $k$ is connected to $t$ by
\begin{equation}
k=\sqrt{\frac{1}{2}\left(1+\frac{1}{\sqrt{1+t^2}}\right)}.
\end{equation}
In the special case $t=0$: $k=1$, $i_1(0)=\infty$, $j_1(0)=2/3$.
In higher orders one needs the generalizations of the integrals
(\ref{eq:K1def}) and (\ref{eq:L1def}). For $n=3,5,\ldots$
let us consider the momentum integrals
\begin{equation}
K_n=\int\frac{d^3 p}{(2\pi\hbar)^3}\frac{1}{E^n} 
=\frac{A(\alpha)}{\alpha^n}i_n(t)|_{t=\Delta/\alpha}, \quad n>1,
\end{equation}
\begin{equation}
L_n=\int\frac{d^3 p}{(2\pi\hbar)^3}\frac{\nu}{E^n}=
\frac{A(\alpha)}{\alpha^{n-1}}j_n(t)|_{t=\Delta/\alpha}, \quad n>1.
\end{equation}
Here the new dimensionless integrals $i_n(t)$, $j_n(t)$ are
defined for odd $n$ as
\begin{equation}
i_n(t)=\int^\infty_0 x^2 d x\,
\left(\frac{1}{\sqrt{(x^2-1)^2+t^2}}\right)^n, \quad n>1,
\label{eq:i_n_def}
\end{equation}
\begin{equation}
j_n(t)=\int^\infty_0 x^2 d  x
\frac{x^2-1}
{\left(\sqrt{(x^2-1)^2+t^2}\right)^n},\quad n>1,
\label{eq:j_n_def}
\end{equation}
For similar integrals written in a different way see
Ref. \cite{Marini98}). They can obtained analitically from 
$i_1(t)$ and $j_1(t)$ using the rules
\begin{equation}
i_{2n+1}(t)=(-1)^n\frac{1}{1\cdot3\cdots(2n-1)}
\left(\frac{1}{t}\frac{\partial}{\partial t}\right)^ni_1(t)
\end{equation}
\begin{equation}
j_{2n+1}(t)=(-1)^{n-1}\frac{1}{1\cdot3\cdots(2n-1)}
\left(\frac{1}{t}\frac{\partial}{\partial t}\right)^nj_1(t),
\end{equation}
which can be easily seen from definitions (\ref{eq:i_n_def}) and
(\ref{eq:j_n_def}) respectivelly. Useful properties performing the
gradient expansions are
\begin{eqnarray}
j'_1(t)&=& t\, i_3(t), \\
i'_n(t)&=&-n\, t\, i_{n+2}(t),\\
j'_n(t)&=&-n\, t\, j_{n+2}(t), \quad n>1.
\end{eqnarray}
In calculating explicitly $i_n(t)$ and $j_n(t)$ for odd $n$
using the
well known formuli for the derivatives of complete elliptic functions
\cite{GradsteinRyzhick} it turns out that they are linear
combinations of $i_1(t)$ and $j_1(t)$:
\begin{eqnarray}
i_n(t)&=&A_n(t)i_1(t)+B_n(t)j_1(t) \\ 
j_n(t)&=&C_n(t)i_1(t)+D_n(t)j_1(t),
\end{eqnarray}
where the coefficients $A_n(t)$, $B_n(t)$, $C_n(t)$ and $D_n(t)$
are rational functions of $t$.

\section{Second order coefficients}
\label{sec:functions}

\begin{figure}
\centerline{
\epsfig{file=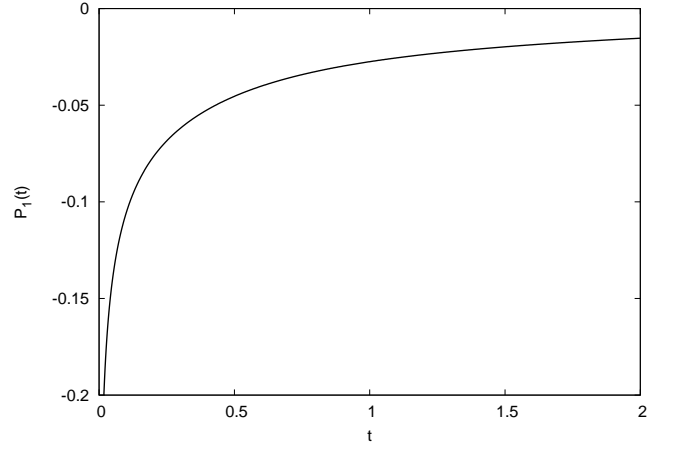,height=\hsize,angle=270}}
\caption{The $P_1(t)$ function (see Eq. (\ref{eq:p1})). \label{fig:p1t}}
\end{figure}

Here we enumerate some dimensionless functions used in the main text. 
Functions occuring in Eq. (\ref{eq:n2h}) are
\begin{eqnarray}
48H_1(t)&=&8\,t^2\,i_5(t)-10\,t^4\,i_7(t)
-t^2\,j_5(t)-10\,t^4\,j_7(t),\nonumber\\
8H_2(t)&=&5\,t^3\,j_7(t)-2\,t\,j_5(t),\nonumber\\
48H_3(t)&=&2i_3(t)-17\,t^2\,i_5(t)
+15\,t^4\,i_7(t)+2j_5(t),\nonumber\\
48H_4(t)&=&2\,t\,i_3(t)+5\,
t^3\,i_5(t)-10\,t^5\,i_7(t)\nonumber\\
&&-4\,t\,j_5(t)+10\,t^3\,j_7(t),\nonumber\\
16H_5(t)&=&4\,t^2\,i_5(t)-5\,t^4\,i_7(t),
\nonumber\\
\label{eq:hak}
\end{eqnarray}
and those used in (\ref{eq:mu2M}) are
\begin{eqnarray}
48M_1(t)&=&10\,t^3\,j_7(t)-4\,t\,j_5(t)
+2\,t\,i_3(t)\nonumber\\
&&+5\,t^3\,i_5(t)-
10\,t^5\,i_7(t),\nonumber\\
8M_2(t)&=&i_3(t)-5\,t^2\,i_5(t)+5\,t^4\,i_7(t),
\nonumber\\
48M_3(t)&=&11\,t\,j_5(t)-15\,t^3\,j_7(t)
+2\,t\,i_5(t),\nonumber\\
48M_4(t)&=&-3\,j_3(t)-t^2\,j_5(t)+10\,t^4\,j_7(t)\nonumber\\
&&-10\,t^2\,i_5(t)+10\,t^4i_7(t),\nonumber\\
16M_5(t)&=&5\,t^3\,j_7(t)-2\,t\,j_5(t).\nonumber\\
\label{eq:mek}
\end{eqnarray}

Straightforward, but lengthy calculation leads to
the analytic forms of the coefficient functions $P_1(t),Q_1(t)$ occuring first
in Eq. (\ref{eq:b2font}):
\begin{gather}
P_1(t)=-\frac{\left[\,8+3\,t^2\right]\,i_1(t)}{384(1+t^2)}
-\frac{5\,j_1(t)}{128(1+t^2)} \nonumber\\
 -\frac{t^4\left[\,1+t^2\right]\,i^4_1(t)}
{192\left(3\,j_1(t)-t^2\,i_1(t)\right)^3}-
\frac{t^2\,i^3_1(t)\left[\,t^2+3\,\right]}
{192\left(3j_1(t)-t^2\,i_1(t)\right)^2}\nonumber\\
 -\frac{i^2_1(t)\left[\,4+3\,t^2\right]}
{384\left(3j_1(t)-t^2\,i_1(t)\right)},
\label{eq:p1}
\end{gather}
\begin{figure}
\centerline{
\epsfig{file=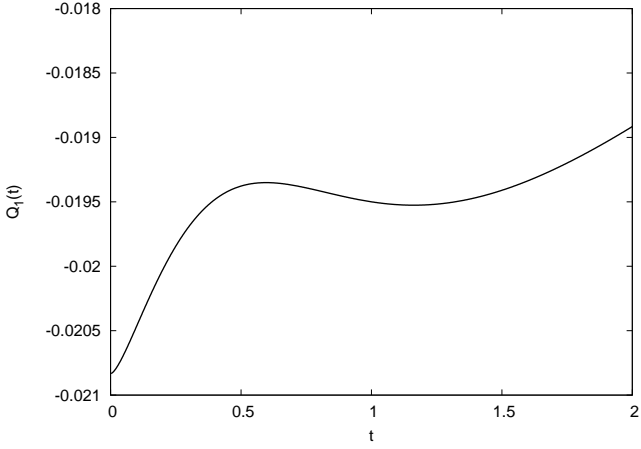,height=\hsize,angle=270}}
\caption{The $Q_1(t)$ function (see Eq. (\ref{eq:q1})).\label{fig:q1t}}
\end{figure}
\begin{equation}
Q_1(t)=\frac{t^2\,i_1(t)}{96(1+t^2)}-
\frac{j_1(t)}{32(1+t^2)}+\frac{t^2\,i^2_1(t)}{96\left(\,3j_1(t)-t^2\,i_1(t)
\right)}.
\label{eq:q1}
\end{equation}
\begin{figure}
\centerline{
\epsfig{file=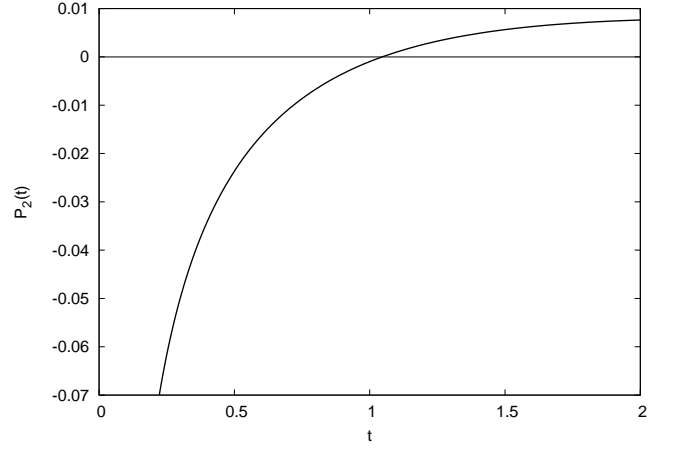,height=\hsize,angle=270}}
\caption{The $P_2(t)$ function (see Eq. (\ref{eq:p2})).\label{fig:p2t}}
\end{figure}
\begin{figure}[t]
\centerline{
\epsfig{file=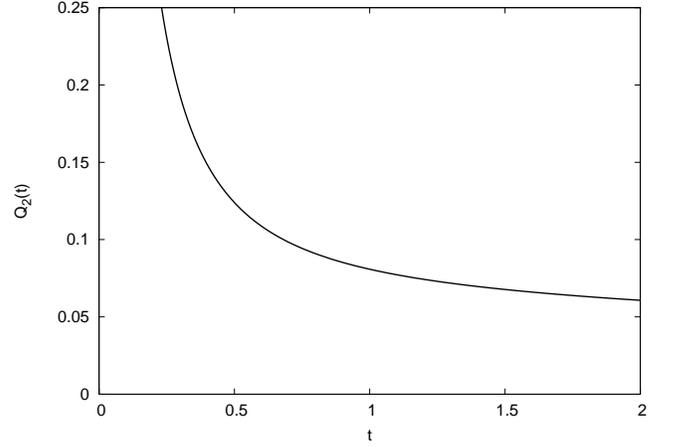,height=\hsize,angle=270}}
\caption{The $Q_2(t)$ function (see Eq. (\ref{eq:q2})).\label{fig:q2t}}
\end{figure}

Similar calculation gives the expressions for $P_2(t),Q_2(t)$ (used in Eq. 
(\ref{eq:d2font})):
\begin{gather}
P_2(t)=-\frac{j_1(t)}{48\,t}+\frac{5\,t
\left[\,i_1(t)+3j_1(t)\right]}{384(1+t^2)}\nonumber\\
-\frac{t\,i_1(t)\left[\,5i_1(t)+7j_1(t)\right]}
{384\left(\,t^2\,i_1(t)-3j_1(t)\right)}\nonumber\\
+\frac{t\,i_1(t)\left[-8i^2_1(t)+i_1(t)j_1(t)+21j^2_1(t)\right]}
{384\left(t^2\,i_1(t)-3j_1(t)\right)^2}\nonumber\\
+\frac{\left[\,i_1(t)+3j_1(t)\right]\,t\,i_1(t)j_1(t)
\left[\,10i_1(t)+21j_1(t)\right]}{192\left(\,t^2\,i_1(t)-3j_1(t)\right)^3},
\label{eq:p2}
\end{gather}
\begin{gather}
Q_2(t)=\frac{i_1(t)+3j_1(t)}{72\,t}+
\frac{t\,\left[\,i_1(t)+3j_1(t)\right]}{96(1+t^2)}\nonumber\\
-\frac{t\,i_1(t)\left[\,10\,i_1(t)+21j_1(t)\right]}
{288\left(t^2i_1(t)-3j_1(t)\right)}.
\label{eq:q2}
\end{gather}
The functions $P_1(t)$, $Q_1(t)$, $P_2(t)$ and $Q_2(t)$ are shown in
Figs.~\ref{fig:p1t}, \ref{fig:q1t}, \ref{fig:p2t} and \ref{fig:q2t} respectively.
At the Feshbach resonance $i_1(T)=0$ should be taken. In this case
\begin{equation}
P_1(T)=-\frac{5 j_1(T)}{128(1+T^2)},\quad
Q_1(T)=-\frac{j_1(T)}{32(1+T^2)},
\label{eq:specp1}
\end{equation} 
\begin{equation}
P_2(T)=-\frac{j_1(T)}{48T}+\frac{5 T j_1(T)}{128(1+T^2)},
\end{equation}
\begin{equation}
Q_2(T)=\frac{j_1(T)}{24T}+\frac{ T j_1(T)}{32(1+T^2)}.
\label{eq:specq2}
\end{equation}

\end{document}